# Noise Measurement of Interacting Ferromagnetic Particles with High Resolution Hall Microprobes

K. Komatsu[a], D. L'Hôte[a], S. Nakamae[a], F. Ladieu[a], V. Mosser[b], A. Kerlain[b], M. Konczykowski[c], E. Dubois[d], V. Dupuis[d], and R. Perzynski[d]

[a] *Service de Physique de l'Etat Condensé (CNRS/MIPPU/URA 2464), DSM/IRAMIS/SPEC, CEA Saclay, F-91191 Gif/Yvette Cedex, France*
[b] *ITRON SAS, 76 avenue Pierre Brossolette, F-92240 Malakoff, France*
[c] *Laboratoire des Solides Irradiés, Ecole Polytechnique, F-91128 Palaiseau, France*
[d] *Laboratoire des Liquides Ioniques et Interfaces Chargées, UMR 7612 CNRS, Université Pierre et Marie Curie, - 4 place Jussieu, Boîte 51, 75252 Paris Cedex 05, France*

**Abstract.** We present our first experimental determination of the magnetic noise of a superspin-glass made of < *1 pico-liter* frozen ferrofluid. The measurements were performed with a local magnetic field sensor based on Hall microprobes operated with the spinning current technique. The results obtained, though preliminary, qualitatively agree with the theoretical predictions of Fluctuation-Dissipation theorem (FDT) violation [1].

## INTRODUCTION

One of the most actively studied areas in the physics of complex systems such as spin glasses, polymers and colloids, is the dynamic correlation length that develops among interacting elements (spins, electrons, molecules, *etc.*) [2]. These length scales manifest themselves as various dynamically heterogeneous phenomena *e.g.,* aging and critical behavior close to a phase transition. Recently, it has been shown theoretically that the mesoscopic out-of-equilibrium fluctuations (noise) and their relation to dissipation (fluctuation-dissipation (FD) relation) should reveal previously unknown spatial heterogeneity of the system [3] including the dynamic correlation length scales. However, there is little or no experimental reports on such *local* spatial-temporal correlation noise due to the extreme weakness of the thermodynamic fluctuations involved [4]. Indeed, the only trusted measurements on fluctuations in out-of-equilibrium systems were performed on 'bulk' samples, *e.g.* in spin-glasses [5] and in structural glasses [6], where the violation of the fluctuation-dissipation theorem (FDT) was observed.

In order to measure the fluctuations at mesoscopic scales, it is desirable to maximize the response from the individual elements of the system as well as the volume occupied by them. For this purpose, a concentrated *ferrofluid superspin glass* is a promising candidate. A 'ferrofluid' consists of ferromagnetic nanoparticles suspended in liquid matrix (in our case, glycerin). When the nanoparticles are sufficiently concentrated, long-range dipolar interactions among them produce spin-glass like behavior (aging, memory, etc.) at low temperatures. These systems are called 'superspin glasses' [7]. Due to their large magnetic moments, the magnetic fluctuations of a superspin glass can become accessible by a micro-meter sized high resolution magnetic field probes placed within a close vicinity of the sample [8].

Here, we report the first successful experimental attempts to measure local magnetic noise (micrometer scale) in 0.5 pico-liter of ferrofluid using a high resolution micro-Hall probe with spinning current technique [9].

## EXPERIMENTAL

Hall microprobes used in our study are Quantum Well Hall Sensors (QWHS) based on (pseudomorphic) AlGaAs/InGaAs/GaAs heterostructures [9, 10, 11]. Such carrier confinement provides a temperature independent carrier density over a wide temperature range ($4 < T < 350$K). The effective field sensitive Hall cross area of $1.6\times1.6\ \mu m^2$ (Fig. 1 left panel) is located at 650 nm beneath the sensor surface. A small, 0.5 - 1 pl drop of ferrofluid made of $\gamma$-$Fe_2O_3$ maghemite nanoparticles (diameter ≈ 8.6nm, magnetic moment ≈ $10^4\ \mu_B$) dispersed in glycerin (volume fraction ≈15%) is deposited directly on the probe surface as seen in Fig. 1, right panel. At low temperatures, the fluid (glycerin) is frozen and the only remaining magnetic degree of freedom is that of the particle magnetic moments. These moments (superspins) interact through the dipolar interaction leading to a superspin-glass transition at $T_g \approx$ 69.5K. Detailed ferrofluid characteristics and preparation methods are found in [12]. In our previous attempts, the Hall microprobes' field resolution was limited to $10\sim20\times10^{-7}$ T for $40 < T < 90$ K, not sufficient to measure the magnetization fluctuations of a superspin-glass. Recently Kerlain *et al.*, reported significant field resolution improvements on 2DEG based Hall microprobes [9, 11] using the *spinning current* technique [13]. We have employed the same approach and achieved so far, a 10 fold field resolution improvement at 77K.

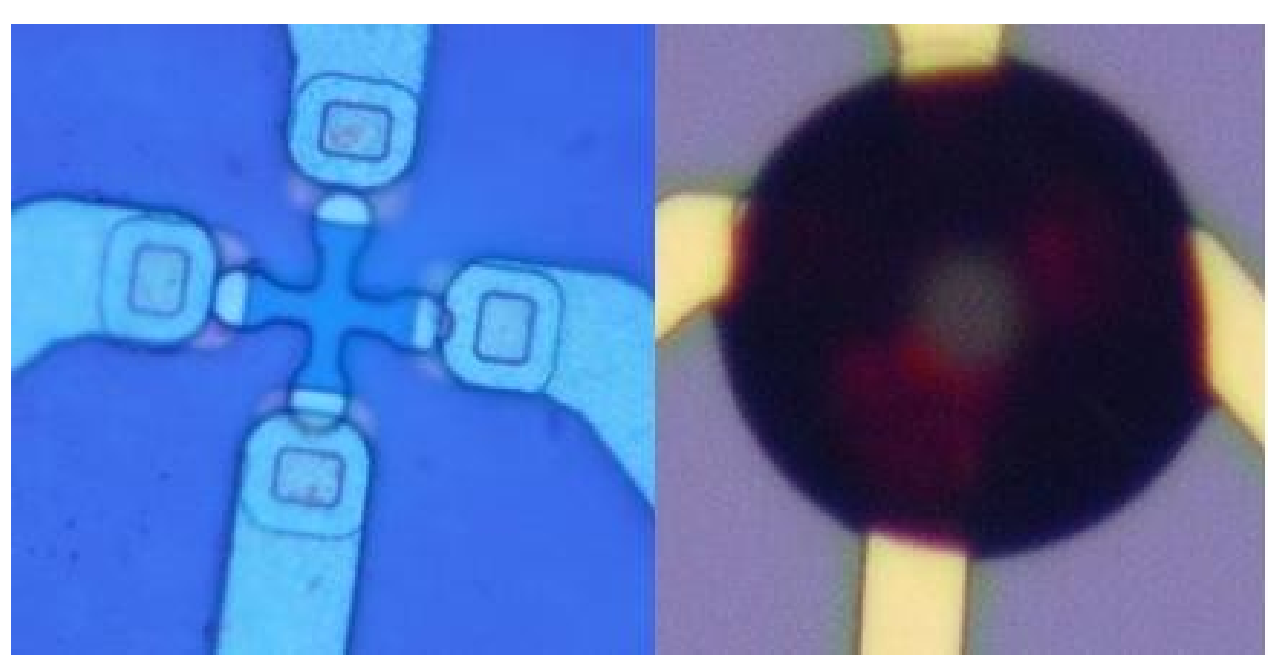

**FIGURE 1.** Left Panel: A Hall microprobe with a nominal cross surface area of $2x2\mu m^2$. Right panel : The same microprobe with 0.5~1.0 pl (12-15 μm in diameter)of $\gamma$-$Fe_2O_3$ ferrofluid deposited on top.

## RESULTS AND DISCUSSION

Prior to measuring the magnetic fluctuation signal of ferrofluid, we characterized the Hall voltage noise spectra of a pristine Hall probe at various temperatures and applied fields. Subsequently, the ferrofluid was deposited on the probe, and cooled down to temperatures ranging from 4.2 to 80 K. Figure 2a shows the spectral magnetic field noise density ($S_B$) as a function of frequency, $f$, at 60K in zero applied field with and without the ferrofluid. The spectra are rather noisy, due to the short data acquisition time (~10 minutes). This is because at 60 K, the ferrofluid is in the superspin glass state, therefore, it is necessary to perform the magnetization noise measurements 'before' the system reaches its equilibrium state. The magnetic noise of ~ 2 mG becomes apparent at frequencies below 3Hz. In the inset of Figure 2a, the squared magnetic noise ($\Delta S_B^2$) due to fluctuations in the frozen ferrofluid sample at 60 and 77 K are depicted. $\Delta S_B$ was estimated by subtracting the background noise of the pristine Hall sensor (black curve in Figure 2a, fitted to a power law function). According to FDT, the quantity $\Delta S_B^2$ is related to the out-of-phase magnetic susceptibility $\chi''(f)$ via $\Delta S_B^2 \sim \chi''(f)/f$ . In order to test the applicability of FDT, we have estimated the $f$-dependence form of $\chi''(f)/f$ from the ac-susceptibility data taken on a bulk sample (~1.5 μl, with the same concentration), as shown in Figure 2b. $\chi''(f)$ was found to increase as $\sim f^{0.127}$ at 77 K and nearly $f$-independent at 60 K for the frequency range of our interest (below 10 Hz). Therefore, if FD relation is obeyed, one would expect to obtain $\Delta S_B^2$ to behave as $\sim 1/f^{0.87}$ and $1/f$ at 77 and 60K, respectively. As can be seen from the inset in Figure 2a, at 77 K where the sample is in the paramagnetic state, $\Delta S_B^2$ follows the f-dependence predicted by FDT (the solid red line) as expected. At 60 K, however, $\Delta S_B^2$ deviates from FDT predicted $1/f$ tendency at frequencies below 1 Hz, hinting at a possible sign of FDT violation in the out-of-equilibrium state.

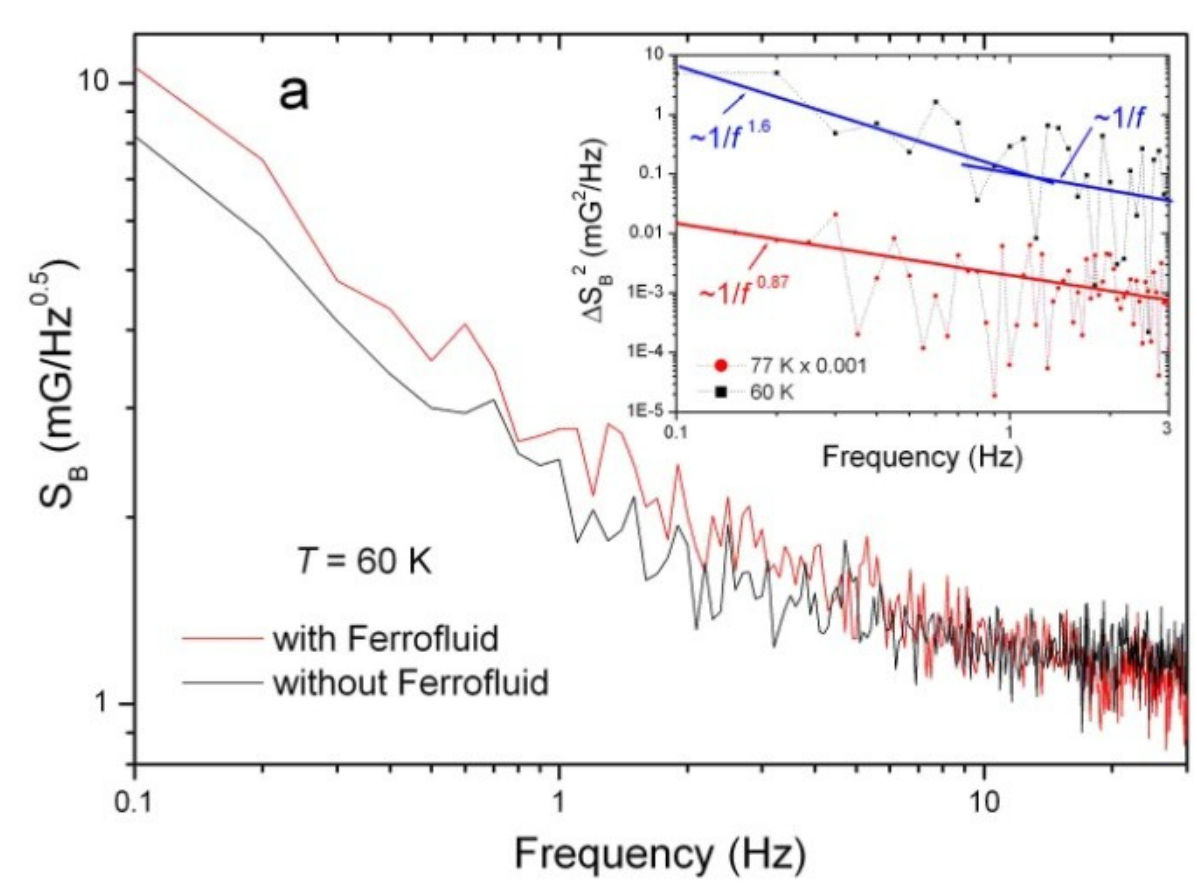

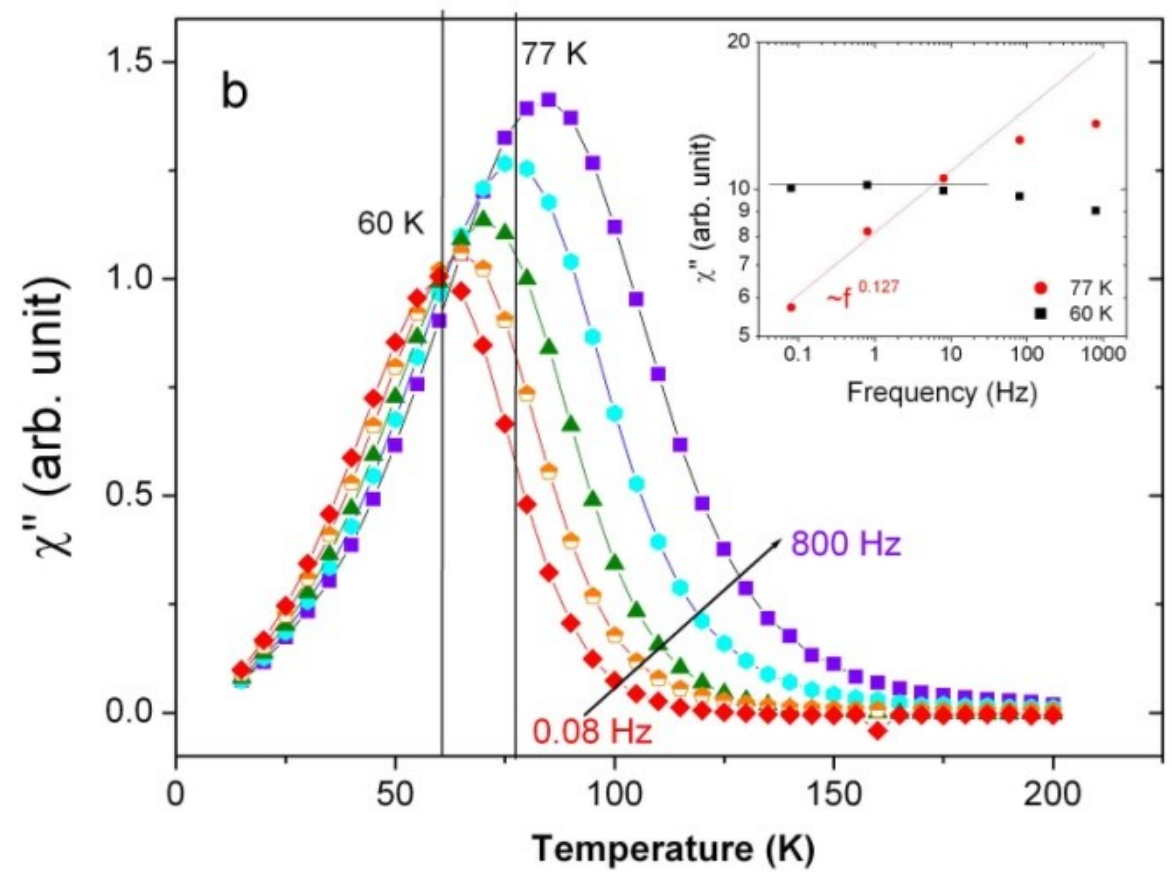

**FIGURE 2.** Left panel: Spectral magnetic field noise density as a function of frequency at 60 K in zero applied field with and without ferrofluid. The magnetic noise becomes apparent at $f < 3$Hz. The inset shows the squared magnetic noise due to ferrofluid magnetization fluctuation (see text) as a function of frequency. Right Panel: The out of phase component of ac magnetic susceptibility $\chi$" vs. temperature of a bulk sample. The inset depicts $\chi$"($f$) at 60 and 77 K (see text).

# CONCLUSION

In summary, we have observed the local magnetic noise stemming from a less than ~ 1 pl ferrofluid sample above and below $T_g$ using a high sensitivity Hall microprobe. To the best of our knowledge, this is the first successful experimental attempt to measure the magnetization fluctuation close to the mesoscopic limit in a (super)spin glass system. The experimental results imply that the FDT is obeyed at temperatures above $T_g$, where the system is in equilibrium. Below $T_g$, on the other hand, a sign of FDT violation has been found. Further improvements of the field resolution are under way. Although the results presented here are preliminary, the technique appears to be promising to study the FDT violation below $T_g$ through the noise measurement.

# ACKNOWLEDGMENTS

We thank Roland Tourbot for his help in the realization of the experimental setup. This work is supported by the RTRA-Triangle de la Physique (MicroHall).